\documentclass[showpacs,prc,preprint,nofootinbib,showkeys]{revtex4}
\pdfoutput=1

\usepackage{bm}
\usepackage{amsmath}
\usepackage{graphicx}
\usepackage{subfigure}
\usepackage[colorlinks=true,linktocpage=true,linkcolor=blue,citecolor=blue]{hyperref}
\usepackage[usenames]{color}

\begin{document}

\title[Anisotropic hydrodynamics for massive systems]{ 
Leading-order anisotropic hydrodynamics for systems with massive particles}

\author{Wojciech Florkowski}
\affiliation{The H. Niewodnicza\'nski Institute of Nuclear Physics, Polish Academy of Sciences, PL-31342 Krak\'ow, Poland} 
\affiliation{Institute of Physics, Jan Kochanowski University, PL-25406~Kielce, Poland}

\author{Radoslaw Ryblewski}
\affiliation{Department of Physics, Kent State University, Kent, OH 44242 United States}
\affiliation{The H. Niewodnicza\'nski Institute of Nuclear Physics, Polish Academy of Sciences, PL-31342 Krak\'ow, Poland} 

\author{Michael Strickland} 
\affiliation{Department of Physics, Kent State University, Kent, OH 44242 United States}

\author{Leonardo Tinti} 
\affiliation{Institute of Physics, Jan Kochanowski University, PL-25406~Kielce, Poland}

\begin{abstract}
The framework of anisotropic hydrodynamics is generalized to include finite particle masses.  Two schemes are introduced and their predictions compared with exact solutions of the kinetic equation in the relaxation time approximation. The first formulation uses the zeroth and first moments of the kinetic equation, whereas the second formulation uses the first and second moments.  For the case of one-dimensional boost-invariant expansion, our numerical results indicate that the second formulation yields much better agreement with the exact solutions.
\end{abstract}



\maketitle

\section{Introduction}
\label{sect:intro}

Heavy-ion experimental data from RHIC (Relativistic Heavy-Ion Collidier) and the LHC (Large Hadron Collider) seem to be very well described by second-order viscous hydrodynamics \cite{Israel:1976tn,Israel:1979wp,Muronga:2001zk,
Muronga:2003ta,Heinz:2005bw,
Baier:2006um,Baier:2007ix,Romatschke:2007mq,
Dusling:2007gi,Luzum:2008cw,Song:2008hj,El:2009vj,
PeraltaRamos:2010je,Denicol:2010tr,Denicol:2010xn,
Schenke:2010rr,Schenke:2011tv,Bozek:2009dw,
Bozek:2011wa,Niemi:2011ix,Niemi:2012ry,
Bozek:2012qs,Denicol:2012cn,PeraltaRamos:2012xk,
Jaiswal:2013npa} with early starting times, $\tau_0 \lesssim$ 1 fm/c. However, in practice one finds that viscous corrections combined with rapid longitudinal expansion result in a substantial pressure asymmetry at early times. This is in agreement with expectations from a variety of microscopic models (string models, color glass condensate, pQCD kinetic calculations), which also predict large momentum anisotropies at early times.   In fact, even in the limit of infinitely strong coupling, where the AdS/CFT correspondence can be used as an effective model, one finds a significant difference between the transverse pressure ${\cal P}_T$ and the longitudinal pressure ${\cal P}_L$, which slowly decays with time \cite{Heller:2011ju,Heller:2012je}.  Such large pressure anisotropies are a cause for concern since viscous hydrodynamics is based on a linearization around a momentum-space isotropic background. The presence of very large shear corrections (of the order of the isotropic pressure) violates the implicit assumption of near-equilibrium evolution and may lead to unphysical results such as negative one-particle distribution functions, negative longitudinal pressure, etc.  Normally, such problems are worrisome only near the edges of the nuclear overlap region; however, when one performs event-by-event hydrodynamics simulations, there can be regions dispersed throughout the volume where the viscous hydrodynamics assumptions are violated.

In order to more accurately treat systems that can possess potentially large pressure anisotropies, a new approach called anisotropic hydrodynamics (aHydro) was developed~\cite{Florkowski:2010cf,Martinez:2010sc,Ryblewski:2010bs,Martinez:2010sd,Ryblewski:2011aq,Florkowski:2011jg,Martinez:2012tu,Ryblewski:2012rr,Florkowski:2012as,Florkowski:2013uqa,Bazow:2013ifa,Tinti:2013vba,Florkowski:2014txa}.  This framework is a reorganization of the conventional hydrodynamic expansion.  In aHydro, the pressure anisotropy is explicitly included at leading order of the hydrodynamic expansion by using a momentum-space anisotropic form as the leading term in the non-equilibrium distribution function.  The evolution equations of leading-order anisotropic hydrodynamics are then solved non-perturbatively.  The anisotropic hydrodynamics framework is appealing because (i) it has been shown in the massless case that aHydro agrees with traditional viscous hydrodynamics in the limit of small anisotropies, (ii) it can be used to describe the large viscosity limit in which the system is longitudinally free streaming and possesses very large anisotropies, (iii) it is guaranteed that the leading-order one-particle distribution and longitudinal pressure are positive both at small and large anisotropy.  Recent developments in anisotropic hydrodynamics have followed two main directions: the first is the systematic computation of next-to-leading-order corrections to a spheroidally-symmetric one-particle distribution function~\cite{Bazow:2013ifa} and the second is to generalize the leading-order form to allow for ellipsoidal one-particle distribution functions in the local rest frame~\cite{Tinti:2013vba}.

In this paper we address an open issue in anisotropic hydrodynamics concerning how to describe systems of massive particles within this framework.  Two different moments-based schemes are considered herein. The first formulation uses the zeroth and first moments of the kinetic equation, whereas the second formulation uses the first and second moments. The predictions of the two formulations are compared with exact solutions of the kinetic equation treated in the relaxation time approximation. Our numerical results for a system of massive particles undergoing one-dimensional boost-invariant expansion indicate that the second formulation of anisotropic hydrodynamics yields much better agreement with the exact solution obtained recently in Ref.~\cite{Florkowski:2014sfa}.

Our results provide further evidence that it is better to use the second moment of the kinetic equation in construction of anisotropic hydrodynamics equations. Previously, it was shown that, for massless particles, the dynamical equations obtained from the second moment more straightforwardly guarantee agreement with Israel-Stewart theory when transverse expansion was included~\cite{Tinti:2013vba}.  Additionally, in Ref.~\cite{Florkowski:2014txa}, where a mixture of anisotropic fluids was analyzed, the two zeroth-moment equations were shown to be insufficient to close the system of equations necessary for the independent evolution of two transverse-momentum scales and two anisotropy parameters.  All of this together suggests that one should use an ellipsoidal ansatz with the equation of motion determined using the first and second moments as the leading-order approximation for future development of the anisotropic hydrodynamics framework.\footnote{Of these two, using the second moment seems to be the most important ingredient.  As we will see below, in the (0+1)-d limit, both formulations of anisotropic hydrodynamics result in spheroidal forms and the only fundamental difference remaining is choice of moments used to obtain the equations of motion.}

Our paper is structured as follows: In Sec. \ref{sect:kineq} we introduce the kinetic equation in the relaxation-time approximation, define the vector basis used for tensor decompositions, and present two alternative (in our case equivalent) parameterizations of anisotropic distribution functions.  In Secs.~\ref{sect:zerothmom}, \ref{sect:firstmom}, and \ref{sect:secondmom} the zeroth, first, and second moments of the kinetic equation are obtained, respectively. Our results are presented in Sec.~\ref{sect:res}.  We conclude and give an outlook in Sec.~\ref{sect:concl}.

\section{Kinetic equation}
\label{sect:kineq}

In this Section we discuss the underlying kinetic equation we use.  We then specialize to the case of a transversely homogeneous system undergoing boost-invariant longitudinal expansion.

\subsection{Relaxation time approximation and equilibrium densities}
\label{sect:rta}

In this paper we consider a simple form for the kinetic equation 
\begin{equation}
 p^\mu \partial_\mu  f(x,p) =  C[f(x,p)] \, ,
\label{kineq}
\end{equation}
where $f(x,p)$ is the one-particle distribution function, and $C$ is the collision term treated in the relaxation time approximation,
\begin{eqnarray}
C[f] = p \cdot u \,\, \frac{f_{\rm eq}-f}{\tau_{\rm eq}} \, ,
\label{col-term}
\end{eqnarray}
with $\tau_{\rm eq}$ being the relaxation time and $u^\mu$ being the hydrodynamic flow of matter. For simplicity, the background equilibrium distribution function $f_{\rm eq}$ is taken to be a Boltzmann distribution
\begin{eqnarray}
f_{\rm eq} = \frac{2}{(2\pi)^3} \exp\left(- \frac{p \cdot u}{T} \right).
\label{Boltzmann}
\end{eqnarray}
The factor 2 in Eq.~(\ref{Boltzmann}) accounts for spin degeneracy. As we shall see below, the temperature $T$ is obtained from a dynamical Landau matching condition which demands that the energy density calculated from the non-equilibrium distribution function $f$ is equal to the energy density determined from the equilibrium background $f_{\rm eq}$ at all times. 

In equilibrium, for particles with mass $M$ obeying classical (Boltzmann) statistics one may use the following expressions for particle density, entropy density, energy density, and pressure,
\begin{eqnarray}
n_{\rm eq} &=& \frac{g_0 M^2 T K_2\left( M/T\right)}{\pi^2} \,  , \label{neq} \\
{\cal S}_{\rm eq} &=& \frac{g_0 M^2}{\pi^2} 
 \left[ 4T K_{2}\left( M/T \right) +M K_{1} \left( M/T\right) \right],
\label{sigmaeq} \\
{\cal E}_{\rm eq} &=& \frac{g_0 T M^2}{\pi^2} 
 \left[ 3T K_{2}\left( M/T \right) +M K_{1} \left( M/T \right) \right], 
\label{epsiloneq} \\
{\cal P}_{\rm eq} &=& \frac{g_0 M^2 T^2 K_2\left( M/T \right)}{\pi^2} \, .
\label{Peq}
\end{eqnarray}
Here $g_0$ is the degeneracy factor which counts all internal degrees of freedom except for spin (the spin degeneracy equals 2).

\subsection{Vector basis for tensor decompositions}
\label{sect:vectorbasis}

Following Refs.~\cite{Florkowski:2011jg,Martinez:2012tu,Tinti:2013vba} we introduce a basis of four vectors that can be used to construct all tensor structures necessary in our analysis. For a transversely homogeneous and boost-invariant system, this basis consists of the flow vector
\begin{equation}
u^\mu = (t/\tau,0,0,z/\tau) \,  ,
\label{u}
\end{equation} 
and three other vectors defined by 
\begin{eqnarray}
z^\mu &=& (z/\tau,0,0,t/\tau) \, ,
\label{z} \\
x^\mu &=& (0,1,0,0) \, ,
\label{x} \\
y^\mu &=& (0,0,1,0) \, .
\label{y} 
\end{eqnarray} 
Here $t$ and $z$ are time and space coordinates, and \mbox{$\tau=\sqrt{t^2-z^2}$} is the (longitudinal) proper time.\footnote{The correct meaning of $z$ (spatial coordinate vs. one of the basis vectors) follows from the context.} The operator $\Delta^{\mu\nu}$ which projects on the space orthogonal to $u^\mu$ can be represented as
\begin{equation}
\Delta^{\mu\nu} = g^{\mu\nu}-u^\mu u^\nu = -x^\mu x^\nu - y^\mu y^\nu -z^\mu z^\nu \, .
\label{Delta}
\end{equation}
Using Eqs.~(\ref{u})--(\ref{y}), the Boltzmann distribution (\ref{Boltzmann}) may be rewritten in the equivalent form as
\begin{equation}
f_{\rm eq} = \frac{2}{(2\pi)^3} \exp\left(- \frac{1}{T}\sqrt{(p\cdot x)^2 + (p\cdot y)^2 + (p\cdot z)^2+M^2}\,\, \right).
\label{BoltzmannXYZ}
\end{equation}

\subsection{Anisotropic distribution functions}
\label{sect:anisodistr}

Within the original anisotropic-hydrodynamics approach one assumes that the distribution function $f$ is of Romatschke-Strickland form \cite{Romatschke:2003ms}
\begin{eqnarray}
f &=&  \frac{2}{(2\pi)^3} \exp\left(- \frac{1}{\Lambda}\sqrt{(p\cdot u)^2 + \xi (p\cdot z)^2}\, \right) \nonumber \\
&=&   \frac{2}{(2\pi)^3} \exp\left(- \frac{1}{\Lambda}\sqrt{(p\cdot x)^2 + (p\cdot y)^2 + (1+\xi) (p\cdot z)^2+M^2}\, \right).
\label{RS1}
\end{eqnarray}
To change from the first to the second line in (\ref{RS1}) we use (\ref{Delta}) and the mass-shell condition $p^2 = M^2$. The parameter $\Lambda$ in (\ref{RS1}) defines a typical transverse-momentum scale in the system, while $\xi$ is the anisotropy parameter. We note that spatial part of the vector $z^\mu$ defines the beam axis in this case.
 
In Refs.~\cite{Martinez:2012tu,Tinti:2013vba} a generalized parameterization of the anisotropic distribution function was proposed, namely
\begin{eqnarray}
 f =  \frac{2}{(2\pi)^3} \exp\left(- \frac{1}{\lambda}\sqrt{(1+\xi_x) (p\cdot x)^2 + (1+\xi_y) (p\cdot y)^2 + (1+\xi_z) (p\cdot z)^2+m^2}\, \right). 
\label{TF1}
\end{eqnarray}
The parameterization (\ref{TF1}) becomes important for practical applications where radial flow is present \cite{Tinti:2013vba}, since in such cases the pressure anisotropies in the $x$ and $y$ directions are generally different, which is not included in (\ref{RS1}).  In this work we consider a boost-invariant and transversely homogeneous system in which case the two formulations (\ref{RS1}) and (\ref{TF1}) are completely equivalent; 
however, the form (\ref{TF1}) has some advantages when one uses the second-moment to obtain the 
equations of motion.  We, therefore, present (\ref{TF1}) in its general form as our starting point.

The anisotropy parameters $\xi_i$ in (\ref{TF1}) satisfy the condition \cite{Tinti:2013vba}
\begin{eqnarray}
\sum_i \xi_i = \xi_x + \xi_y + \xi_z = 0 \, .
\label{sumofxis}
\end{eqnarray}
For the case of one-dimensional boost-invariant expansion considered in this paper, the parameterizations (\ref{RS1}) and (\ref{TF1}) are connected through the following set of simple transformations:~\footnote{Note that a constant mass parameter $M$ implies a space-time dependent $m$.}
\begin{eqnarray}
\xi_x &=& -\frac{\xi/3}{1+\xi/3} \,  , \quad
\xi_y =  -\frac{\xi/3}{1+\xi/3} \,  , \quad
\xi_z = \frac{2\,\xi/3}{1+\xi/3}\,  , \nonumber \\
\lambda &=& \Lambda (1+\xi/3)^{-1/2} \, , \quad
m = M (1+\xi/3)^{-1/2} \,  .
\label{RS-TF}
\end{eqnarray}
It is also useful to note that
\begin{eqnarray}
\sqrt{\Pi_j (1+\xi_j)} = \frac{(1+\xi)^{1/2}}{(1+\xi/3)^{3/2}} \, .
\label{sqrtprod}
\end{eqnarray}

\section{Zeroth moment of the kinetic equation}
\label{sect:zerothmom}

The covariant integration of the kinetic equation (\ref{kineq}) over three-momentum gives 
\begin{eqnarray}
\partial_\mu N^\mu = u_\mu \frac{N^\mu_{\rm eq}-N^\mu}{\tau_{\rm eq}},
\label{zmom1}
\end{eqnarray}
where the particle number currents $N^\mu_{\rm eq}$ and $N^\mu$ are defined by the equations
\begin{eqnarray}
N^\mu_{\rm eq} = g_0 \int dP \, p^\mu f_{\rm eq} \, ,  \quad
N^\mu = g_0 \int dP \, p^\mu f \, . 
\label{Ns}
\end{eqnarray}
Here we use the shorthand notation $dP$ for the Lorentz-invariant integration measure $dP=d^3p/E_p$ with the particle's energy $E_p=p^0=\sqrt{M^2+p_x^2+p_y^2+p_z^2}$. The form of the equilibrium distribution function (\ref{Boltzmann}) implies that
\begin{eqnarray}
N^\mu_{\rm eq} = n_{\rm eq} u^\mu \, ,
\label{neq1}
\end{eqnarray}
where 
\begin{eqnarray}
n_{\rm eq} = \frac{g_0}{4\pi^3} \int \! dP \;  p \cdot u \, \exp\left(- \frac{p \cdot u}{T} \right).
\label{neq2}
\end{eqnarray}
The integral on the right-hand side of Eq.~(\ref{neq2}) can be performed most easily in the local rest frame of the fluid element where $u^\mu=(1,0,0,0)$. This leads us directly to Eq.~(\ref{neq}).

Repeating the same steps for the anisotropic distribution function $f$ given by Eq.~(\ref{RS1}) we obtain
\begin{eqnarray}
N^\mu = n u^\mu \, ,
\label{nAH1}
\end{eqnarray}
where 
\begin{eqnarray}
n = \frac{g_0}{4\pi^3} \int \! dP \;  p \cdot u \,  \exp\left[-\frac{1}{\Lambda}
\sqrt{(p \cdot u)^2 + \xi (p \cdot z)^2} \right] .
\label{nAH2}
\end{eqnarray}
Again, the integral in Eq.~(\ref{nAH2}) can be performed most easily in the local rest frame of the fluid element where $u^\mu=(1,0,0,0)$ and, in addition, $z^\mu = (0,0,0,1)$. In this way one finds
\begin{eqnarray}
n = \frac{g_0 M^2 \Lambda K_2(M/\Lambda)}{\pi^2 \sqrt{1+\xi}} \, .
\label{nAH3}
\end{eqnarray}
As expected, for isotropic systems one has $\xi=0$ and $\Lambda=T$ and Eq.~(\ref{nAH3}) reduces to Eq.~(\ref{neq}).

Substituting Eqs.~(\ref{neq1}) and (\ref{nAH1}) into
Eq.~(\ref{zmom1}) we find
\begin{eqnarray}
\partial_\mu (n u^\mu) =
\frac{n_{\rm eq}-n}{\tau_{\rm eq}} \, .
\label{zmom2}
\end{eqnarray}
For one-dimensional boost-invariant expansion, the divergence of the flow and the total time derivative are given by simple expressions,
$\partial_\mu u^\mu = 1/\tau$ and $u^\mu \partial_\mu = d/d\tau$, therefore, Eq.~(\ref{zmom2}) leads to
\begin{eqnarray}
\frac{d}{d\tau} \ln n +\frac{1}{\tau}
= \frac{1}{\tau_{\rm eq}}
\left( \frac{n_{\rm eq}}{n}- 1 \right) ,
\label{zmom3}
\end{eqnarray}
or, written explicitly in terms of the microscopic parameters
\begin{eqnarray}
&& \frac{d}{d\tau} \left(
 \ln (1+\xi)^{-1/2} +\ln \Lambda K_2(M/\Lambda)
 \right) +\frac{1}{\tau} = 
 \frac{1}{\tau_{\rm eq}}
\left( \frac{T K_2(M/T)}{\Lambda K_2(M/\Lambda)}
\sqrt{1+\xi}- 1 \right). 
\label{zmom4}
\end{eqnarray}
From (\ref{zmom4}) we obtain 
\begin{eqnarray}
&& -\frac{1}{2(1+\xi)}\frac{d\xi}{d\tau} 
+\left(3 + \frac{M}{\Lambda} \frac{K_1(M/\Lambda)}{K_2(M/\Lambda)} \right) \frac{d\Lambda}{\Lambda d\tau} 
 +\frac{1}{\tau} = 
  \frac{r}{\tau_{\rm eq}}
\left( \frac{T K_2(M/T)}{\Lambda K_2(M/\Lambda)}
\sqrt{1+\xi}- 1 \right). \hspace{5mm}
\label{AHM1}
\end{eqnarray}
On the right-hand side of (\ref{zmom4}) we have introduced a factor $r$. Our previous studies \cite{Florkowski:2013lya,Florkowski:2013lza}  showed that in the massless limit one should take $r = 2 \Lambda/T$ in (\ref{zmom4}) in order to improve agreement between the traditional formulation of anisotropic hydrodynamics and exact solutions of the kinetic equation.  We will use the same prescription for $M\neq0$.  In the limit $M \to 0$, Eq.~(\ref{AHM1}) becomes 
\begin{eqnarray}
&& -\frac{1}{2(1+\xi)}\frac{d\xi}{d\tau} 
+ \frac{3}{\Lambda} \frac{d\Lambda}{ d\tau} 
 +\frac{1}{\tau} = 
  \frac{r}{\tau_{\rm eq}}
\left( \frac{T^3}{\Lambda^3} \sqrt{1+\xi}- 1 \right),
\label{AHM10}
\end{eqnarray}
which is the result originally obtained in Ref.~\cite{Martinez:2010sc}.

\section{First moment of the kinetic equation}
\label{sect:firstmom}

We now proceed to the evaluation of the first moment of the kinetic equation.  We begin by specifying the tensor structure of the energy-momentum tensor for a system of massive particles undergoing one-dimensional boost-invariant expansion.

\subsection{Energy-momentum tensors}
\label{sect:enmomten}

The first moment of the kinetic equation (\ref{kineq}) gives
\begin{eqnarray}
\partial_\mu T^{\mu\nu} = u_\mu 
\frac{T^{\mu \nu}_{\rm eq}-T^{\mu\nu}}{\tau_{\rm eq}} \, .
\label{fmom1}
\end{eqnarray}
Here $T^{\mu \nu}_{\rm eq}$ is the equilibrium energy-momentum tensor  
\begin{eqnarray}
T^{\mu \nu}_{\rm eq} = \left( {\cal E}_{\rm eq}
+ {\cal P}_{\rm eq} \right)  u^\mu u^\nu
- {\cal P}_{\rm eq} g^{\mu\nu}  \, ,
\label{TEQ}
\end{eqnarray}
where ${\cal E}_{\rm eq}$ and ${\cal P}_{\rm eq}$ are given by Eqs.~(\ref{epsiloneq}) and (\ref{Peq}), respectively. If the distribution function $f$ is given by the Romatschke-Strickland form (\ref{RS1}), the energy-momentum tensor $T^{\mu\nu}$ has the structure
\begin{eqnarray}
T^{\mu \nu} &=& \left( {\cal E}
+ {\cal P}_T \right)  u^\mu u^\nu
- {\cal P}_T g^{\mu\nu}
+\left( {\cal P}_L - {\cal P}_T \right) z^\mu z^\nu 
\nonumber \\
&=& {\cal E}\, u^\mu u^\nu
+ {\cal P}_T\, x^\mu x^\nu
+ {\cal P}_T\, y^\mu y^\nu
+ {\cal P}_L\, z^\mu z^\nu  \, .
\label{TAH}
\end{eqnarray}
From the second line of (\ref{TAH}) we see that the two transverse pressures are equal for the case considered herein, however, the system is still anisotropic since the transverse and longitudinal pressures can be different. The energy density ${\cal E}$ appearing in (\ref{TAH}) can be obtained from the formula
\begin{eqnarray}
{\cal E} = \frac{g_0}{4\pi^3} \int dP \,
(p\cdot u)^2 \exp\left[
-\frac{\sqrt{(p\cdot u)^2+\xi (p\cdot z)^2}}{\Lambda}
\right].
\label{epsAH1}
\end{eqnarray}
In order to perform the integral (\ref{epsAH1}) we change to the local rest frame, where
\begin{eqnarray}
{\cal E} =
\frac{g_0}{4\pi^3} \int d^3p
\sqrt{p_L^2 + p_T^2 +M^2} \exp\left[
-\frac{\sqrt{(1+\xi)p_L^2 +p_T^2 +M^2 }}{\Lambda}
\right].
\label{epsAH2}
\end{eqnarray}
By changing the integration variables first to
$a=\sqrt{1+\xi} \, p_L/\Lambda$ and $b=p_T/\Lambda$, and later to $r=\sqrt{a^2+b^2}$ and $\phi = \tan^{-1}(b/a)$, one finds
\begin{eqnarray}
{\cal E} = \frac{g_0 \Lambda^4}{2\pi^2}
\tilde{\cal H}_2\left[(1+\xi)^{-1/2},M/\Lambda \right].
\label{epsAH3}
\end{eqnarray}
The function $\tilde{\cal H}_2$ is defined in the Appendix by Eqs.~(\ref{tildeH2}) and (\ref{H2}).

The calculation of the longitudinal pressure ${\cal P}_L$ in (\ref{TAH}) proceeds in a similar manner. The starting point is
\begin{eqnarray}
{\cal P}_L = \frac{g_0}{4\pi^3} \int dP
(p\cdot z)^2 \exp\left[
-\frac{\sqrt{(p\cdot u)^2+\xi (p\cdot z)^2}}{\Lambda}
\right].
\label{PLAH1}
\end{eqnarray}
Then we change to the local rest frame
\begin{eqnarray}
{\cal P}_L =
\frac{g_0}{4\pi^3} \int \frac{d^3p \,p_L^2}{
\sqrt{p_L^2 + p_T^2 +M^2}} \exp\left[
-\frac{\sqrt{(1+\xi)p_L^2 +p_T^2 +M^2 }}{\Lambda}
\right].
\label{PLAH2}
\end{eqnarray}
By making the same change of the integration variables as in the case of the energy density we obtain
\begin{equation}
{\cal P}_L = \frac{g_0 \Lambda^4}{2\pi^2}
\tilde{\cal H}_{2L} \left[(1+\xi)^{-1/2},M/\Lambda \right] , 
\label{PLAH3}
\end{equation}
where $\tilde{\cal H}_{2L}$ is defined in the Appendix Eq.~(\ref{tildeH2L}). 
Similar steps can be used to calculate the transverse pressure ${\cal P}_T$ with the result being
\begin{equation}
{\cal P}_T = \frac{g_0 \Lambda^4}{4\pi^2}
\tilde{\cal H}_{2T} \left[(1+\xi)^{-1/2},M/\Lambda \right] ,
\label{PTAH3}
\end{equation}
where $\tilde{\cal H}_{2T}$ is defined in the Appendix Eq.~(\ref{tildeH2T}).

\subsection{Dynamical Landau matching condition}
\label{sect:landaumatch}

Since we want to have the energy and momentum conserved in our system, the right-hand side of Eq.~(\ref{fmom1}) should vanish. Using our expressions for the energy-momentum tensors (\ref{TEQ}) and (\ref{TAH}) we find
\begin{eqnarray}
{\cal E} = {\cal E}_{\rm eq} \, ,
\label{LM1}
\end{eqnarray}
or more explicitly
\begin{eqnarray}
\frac{1}{2} \,\Lambda^4
\tilde{\cal H}_{2}\left((1+\xi)^{-1/2},M/\Lambda \right)
=   T M^2
 \left[ 3T K_{2}\left( M/T \right) + M K_{1} \left( M/T \right) \right].
 \label{AHM2}
\end{eqnarray}
Equation (\ref{AHM2}) allows us to express the effective temperature $T$ in terms of the anisotropy parameter $\xi$ and the scale $\Lambda$. In the limit $M \to 0$ one obtains 
\begin{eqnarray}
\frac{1}{2}\,\Lambda^4
{\cal H}\left(\frac{1}{\sqrt{1+\xi}}\right) 
= \Lambda^4 {\cal R}(\xi) = T^4,
 \label{AHM20}
\end{eqnarray}
where the functions ${\cal R}$ and ${\cal H}$ were defined in Refs.~\cite{Martinez:2010sc} and \cite{Florkowski:2013lza}, respectively. Equation (\ref{AHM20}) agrees with the form of the dynamical Landau matching condition introduced for the first time in \cite{Martinez:2010sc}.

\subsection{Energy conservation}
\label{sect:encon}

For one-dimensional boost-invariant systems the four equations included in energy-momentum conservation law $\partial_\mu T^{\mu\nu}=0$ reduce to the single equation
\begin{eqnarray}
\frac{d{\cal E}}{d\tau} =
-\frac{{\cal E}+{\cal P}_L}{\tau} \, ,
\end{eqnarray}
which expresses energy conservation in the system. In our case this leads to the expression
\begin{eqnarray}
&& \frac{d}{d\tau} \left[ \Lambda^4 \tilde{\cal H}_{2}
\left((1+\xi)^{-1/2},M/\Lambda \right) \right] 
 =  \nonumber \\
&&
\hspace{2cm}
-\frac{\Lambda^4}{\tau} \left[
\tilde{\cal H}_{2}
\left((1+\xi)^{-1/2},M/\Lambda \right)
+\tilde{\cal H}_{2L}
\left((1+\xi)^{-1/2},M/\Lambda \right) \right] .
\end{eqnarray}
The properties of the functions $\tilde{\cal H}$ discussed in the Appendix allow us to rewrite this equation as follows
\begin{eqnarray}
\frac{\left[4 \tilde{\cal H}_{2}
\left((1+\xi)^{-1/2},M/\Lambda \right)-\Omega_4(\xi,\Lambda)\right] \, }{\Omega_2(\xi,\Lambda) } \frac{d\Lambda}{\Lambda d\tau}-
\frac{1}{2(1+\xi)} \frac{d\xi}{d\tau}
+\frac{1}{\tau} = 0 \, ,
\label{AHM3} 
\end{eqnarray}
where the functions $\Omega_2$ and $\Omega_4$ are defined in the Appendix.

We are again interested in the case of massless particles. For $M \to 0$ the function $\Omega_4$ vanishes, while $\Omega_2 = 6\left({\cal H}((1+\xi)^{-1/2})+{\cal H}_L((1+\xi)^{-1/2})\right)$, where the functions ${\cal H}$ and ${\cal H}_L$ are defined in \cite{Florkowski:2013lya}. This allows us to rewrite Eq.~(\ref{AHM3}) in the form
\begin{eqnarray}
4 \frac{ {\cal H}}{\Lambda} \frac{d\Lambda}{d\tau}
-\frac{{\cal H}+{\cal H}_L}{2(1+\xi)}\,\,\frac{d\xi}{d\tau} = -\frac{{\cal H}+{\cal H}_L}{\tau},
\end{eqnarray}
where the argument of the functions  ${\cal H}$ and ${\cal H}_L$ is $(1+\xi)^{-1/2}$. Using the connections between the functions ${\cal H}$ and ${\cal R}$ \cite{Florkowski:2013lya} we find a simpler form
\begin{eqnarray}
4  \frac{{\cal R}}{\Lambda} \frac{d\Lambda}{d\tau}
+{\cal R}^\prime\,\,\frac{d\xi}{d\tau} = -\frac{1}{\tau} \left({\cal R}+\frac{1}{3}{\cal R}_L\right),
\label{AHM30}
\end{eqnarray}
which was obtained in earlier formulations of anisotropic hydrodynamics for massless particles, see e.g.~Refs.~\cite{Martinez:2010sc,Florkowski:2012ax}.

\section{Second moment of the kinetic equation}
\label{sect:secondmom}

Now we turn to the discussion of the second moment. Its analysis is performed in a slightly different way,  as we use now the form (\ref{TF1}) of the distribution function since it is more straightforward to work with in this case. This form has turned out to be very convenient for extracting equations of motion for anisotropic hydrodynamics which reproduce Israel-Stewart theory~\cite{Tinti:2013vba} in the situation where the non-zero radial flow is included and the system is close to isotropic equilibrium.
 
\subsection{Tensor decompositions for the second moment}
\label{sect:tendec}

The second moment of the Boltzmann equation may be written in the form analogous to the zeroth and first moments, namely
\begin{eqnarray}
\partial_\lambda \Theta^{\lambda\mu\nu} = \frac{1}{\tau_{\rm eq}} \left(u_\lambda\Theta_{\rm eq}^{\lambda\mu\nu} - u_\lambda\Theta^{\lambda\mu\nu}\right),
 \label{tmom}
\end{eqnarray}
where 
\begin{eqnarray}
\Theta^{\lambda\mu\nu} = g_0 \int\!\! dP \; p^\lambda p^\mu p^\nu f \, , \quad
\Theta^{\lambda\mu\nu}_{\rm eq} = g_0 \int\!\! dP \; p^\lambda p^\mu p^\nu f_{\rm eq} \, .
\label{Thetas}
\end{eqnarray}
For boost-invariant and cylindrically symmetric systems, due to the quadratic dependence of the distribution function on the momentum, the only non-vanishing terms in (\ref{Thetas}) are those with an even number of each spatial index. In the covariant form they read
\begin{eqnarray} 
\Theta &=& \Theta_u \left[ u\otimes u \otimes u\right] 
\nonumber \\
&& \,+\, \Theta_x \left[ u\otimes x \otimes x +x\otimes u \otimes x + x\otimes x \otimes u\right] 
\nonumber \\ 
&& \,+\,  \Theta_y  \left[ u\otimes y \otimes y +y\otimes u \otimes y + y\otimes y \otimes u\right]
\nonumber \\
&& \,+\, \Theta_z \left[ u\otimes z \otimes z +z\otimes u \otimes z + z\otimes z \otimes u\right] .
 \label{Theta}
\end{eqnarray}
The equilibrium tensor has the same decomposition but, due to the rotational invariance of the local equilibrium state, the coefficients $\Theta_{i, \rm eq}\,\,(i=x,y,z)$ are all equal to the term we denote below as $\Theta_{\rm eq}$.

\subsection{Main dynamic equations from the second moment}
\label{sect:maindyneq}

It has been argued in \cite{Tinti:2013vba} that in order to obtain the agreement with the Israel-Stewart theory, the anisotropic-hydrodynamics approach should include the three equations of the form
\begin{eqnarray}
&& \frac{d}{d\tau} \ln\Theta_i + \theta - 2\theta_i -\frac{1}{3}\sum_j \left[\frac{d}{d\tau} \ln\Theta_j + \theta - 2\theta_j \right]  \nonumber \\
&& = \frac{1}{\tau_{\rm eq}} \left[ \frac{\Theta_{\rm eq}}{\Theta_i} - 1 \right] -\frac{1}{3}\sum_j \left\{ \frac{1}{\tau_{\rm eq}} \left[ \frac{\Theta_{\rm eq}}{\Theta_j} - 1 \right]  \right\} \quad (i=x,y,z) \, .
\label{sum}
\end{eqnarray}
These equations follow from (\ref{tmom}) if the following manipulations are performed: first we contract (\ref{tmom}) twice with the vector $i$ with $i \in \{x,y,z\}$, second we divide each of the obtained equations by $\Theta_i$, and finally, from each of the equations obtained, we subtract one-third of their sum. For the one-dimensional boost-invariant case considered herein, the coefficients $\theta_i$ in (\ref{sum}) are $\theta_x=\theta_y=0$ and $\theta_z =-1/\tau$ (with $\theta=\partial_\mu u^\mu = -\theta_x-\theta_y-\theta_z= 1/\tau$).

In the massive case, the coefficients $\Theta_i$ appearing in Eq.~(\ref{Theta}) are given by the following integrals
\begin{equation}
  \Theta_i = g_0 \int\!\! dP \,  (p\cdot u) \, (p\cdot i)^2 f =  \frac{ g_0 m^3 \lambda^2 K_3(m/\lambda) }
  {\pi^2 \sqrt{ \prod_j(1+\xi_j) } } \frac{1}{1+\xi_i} \, ,
\label{Theta_I}
\end{equation}
and
\begin{equation}
  \Theta_{\rm eq} = g_0 \int\!\! dP \,  (p\cdot u) \, (p\cdot i)^2 f_{\rm eq} =  
  \frac{ g_0 M^3 T^2 K_3(M/T) }{\pi^2} \, .
\label{Theta_eq}  
\end{equation}

\subsection{Reduction to the one-dimensional boost-invariant case}
\label{sect:maindyneq0p1}

In our case it is enough to consider only one out of the three equations in (\ref{sum}) --- the projections of (\ref{sum}) with $i=x$ and $i=y$ are identical and the sum of Eqs.~(\ref{sum}) is zero. Therefore, taking the $i=x$ component we find 
\begin{eqnarray}
\frac{1}{1+\xi_x} \frac{d\xi_x}{d\tau} -\frac{1}{3} \sum_j \frac{1}{1+\xi_j} \frac{d\xi_j}{d\tau}
+ \frac{2}{3\tau}  + \frac{\xi_x}{\tau_{\rm eq}}
\frac{M^3 T^2 K_3(M/T)}{m^3 \lambda^2 K_3(M/\Lambda)} \sqrt{\Pi_j (1+\xi_j)} 
= 0 \, ,
\end{eqnarray}
Using Eqs.~(\ref{RS-TF}) and (\ref{sqrtprod}) to switch finally to the Romatschke-Strickland parameterization, one obtains
\begin{eqnarray}
&& -\frac{1}{1+\xi} \frac{d\xi}{d\tau} + \frac{2}{\tau} = \frac{\xi}{\tau_{\rm eq}}
\frac{T^2 K_3(M/T)}{\Lambda^2 K_3(M/\Lambda)} \sqrt{1+\xi} \, ,
\label{AHM4}
\end{eqnarray}
which, in the massless limit, becomes
\begin{eqnarray}
&& -\frac{1}{1+\xi} \frac{d\xi}{d\tau} + \frac{2}{\tau} = \frac{\xi}{\tau_{\rm eq}}
\frac{T^5}{\Lambda^5} \sqrt{1+\xi} \, .
\label{AHM40}
\end{eqnarray}
%

\begin{figure}[t]
\centerline{\includegraphics[angle=0,width=0.7\textwidth]{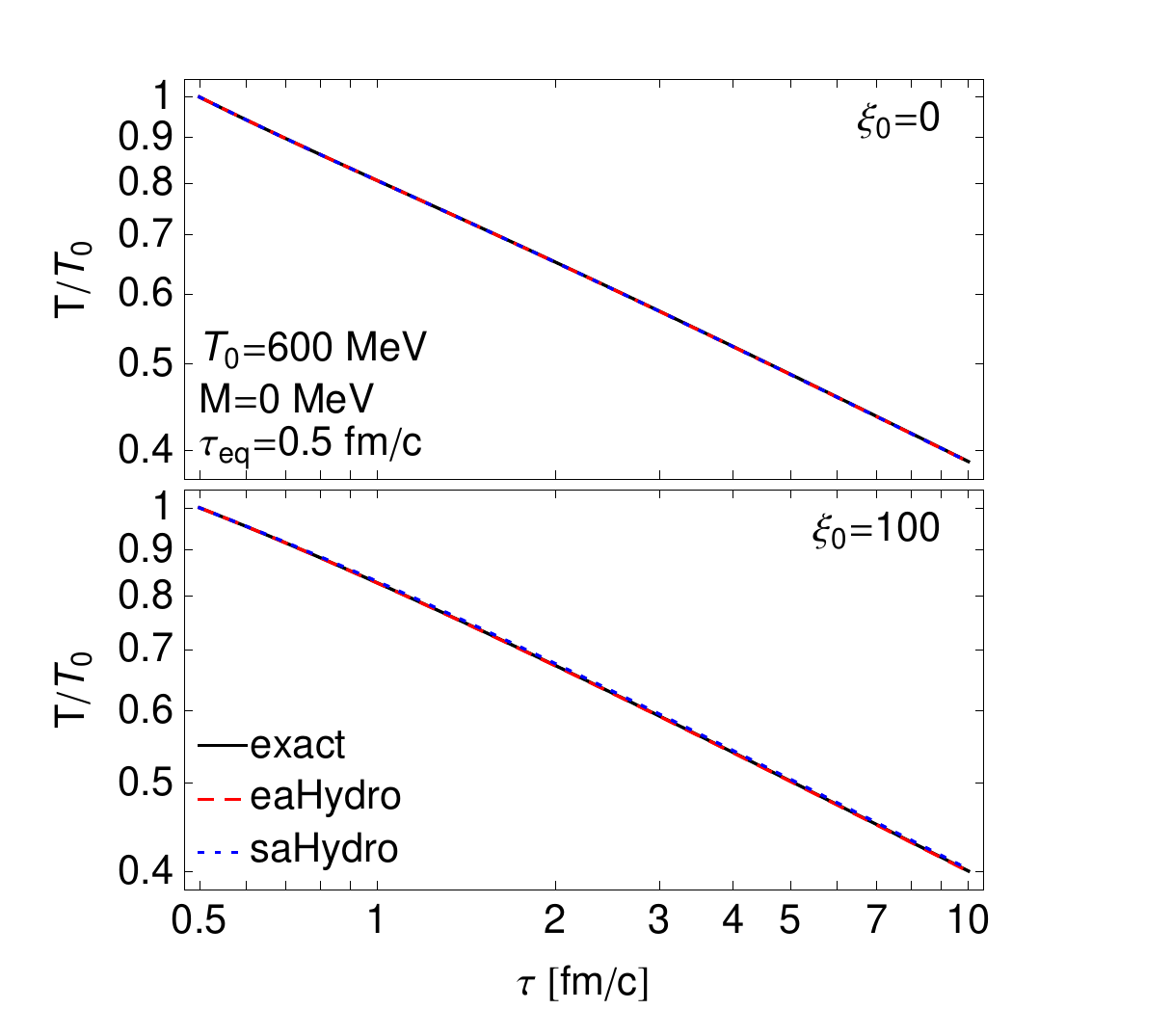}}
\caption{(Color online) Time dependence of the effective temperature.}
\label{fig:Tmass0}
\end{figure}

\begin{figure}[t]
\centerline{\includegraphics[angle=0,width=0.7\textwidth]{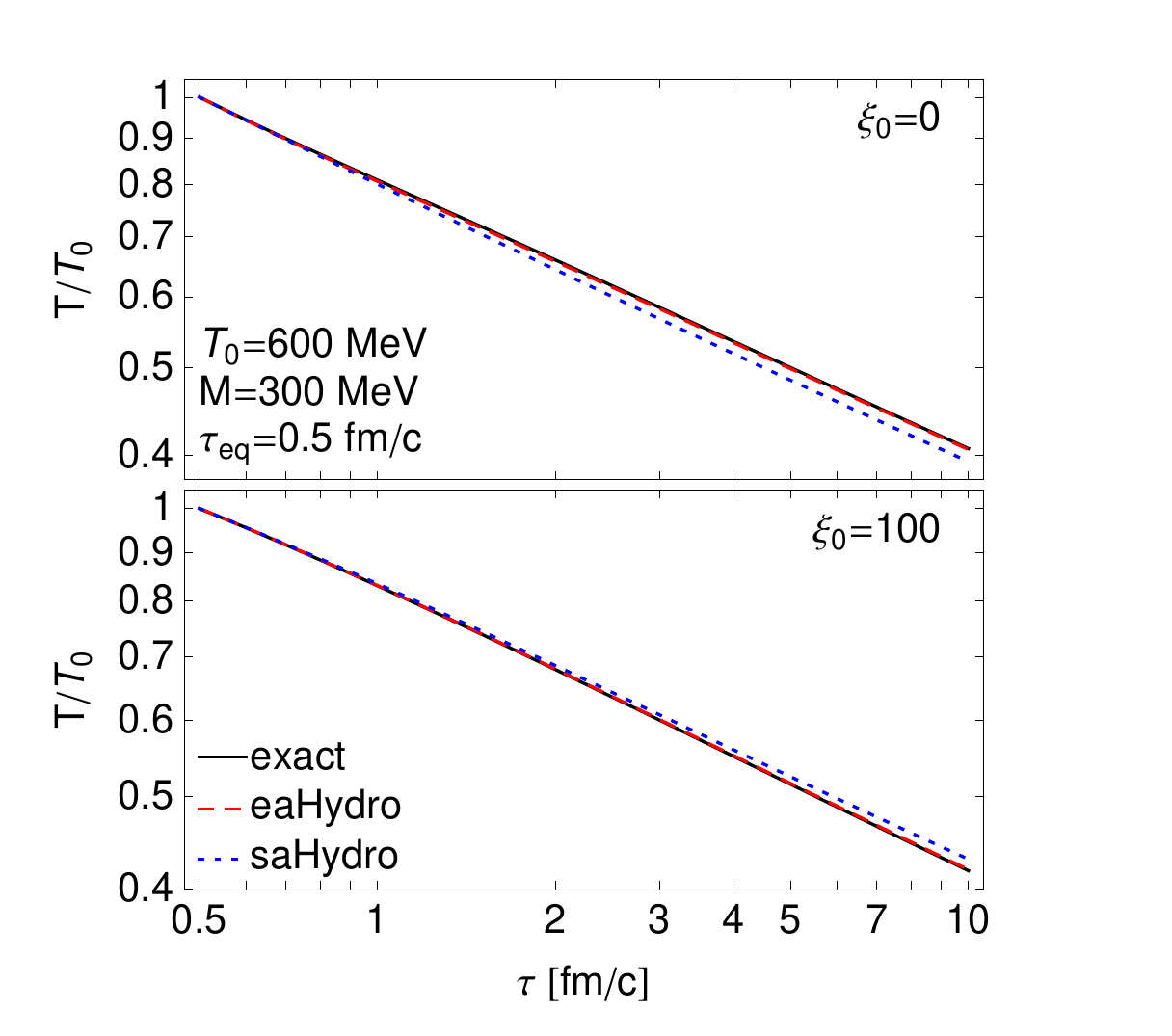}}
\caption{(Color online) Same as Fig.~\ref{fig:Tmass0} but now for the finite particle mass of 300 MeV.}
\label{fig:Tmass}
\end{figure}

\section{Results}
\label{sect:res}

As mentioned previously, in this work we consider two different formulations of anisotropic hydrodynamics in order to assess their relative accuracy. The first approach, which we will call spheroidal anisotropic hydrodynamics (saHydro), uses equations obtained from the zeroth and first moments of the kinetic equation, namely Eqs.~(\ref{AHM1}), (\ref{AHM2}), and (\ref{AHM3}) --- this may be regarded as the traditional approach.  In the second approach, which we will call ellipsoidal anisotropic hydrodynamics (eaHydro), the zeroth-moment equation is replaced by an equation obtained from the second moment, i.e., one uses Eqs.~(\ref{AHM2}), (\ref{AHM3}) and (\ref{AHM4}).  We will also compare the two approaches in the massless limit using Eqs.~(\ref{AHM10}), (\ref{AHM20}) and (\ref{AHM30}) or Eqs.~(\ref{AHM20}), (\ref{AHM30}) and (\ref{AHM40}).  For comparison with the exact solutions published in Ref.~\cite{Florkowski:2014sfa} we take $g_0 = 16$.

We begin with the massless case, $M=0$. In Fig.~\ref{fig:Tmass0} we show the time dependence of the effective temperature $T$ obtained using kinetic theory, saHydro, and eaHydro.  In the top panel the initial conditions used were \mbox{$T_0$ = 600 MeV} and $\xi_0 = 0$, and in the bottom panel we show the case $\xi_0=100$ (highly oblate initial momentum distribution) with the same initial effective temperature.  The solid black line shows the result obtained from the kinetic theory, the dashed red line is the eaHydro result, and the dotted blue line is the saHydro result.  As we can see from this figure, the three results differ very little from one other in this case, demonstrating a generally good agreement between the exact solution and anisotropic hydrodynamics.

\begin{figure}[t]
\centerline{\includegraphics[angle=0,width=0.7\textwidth]{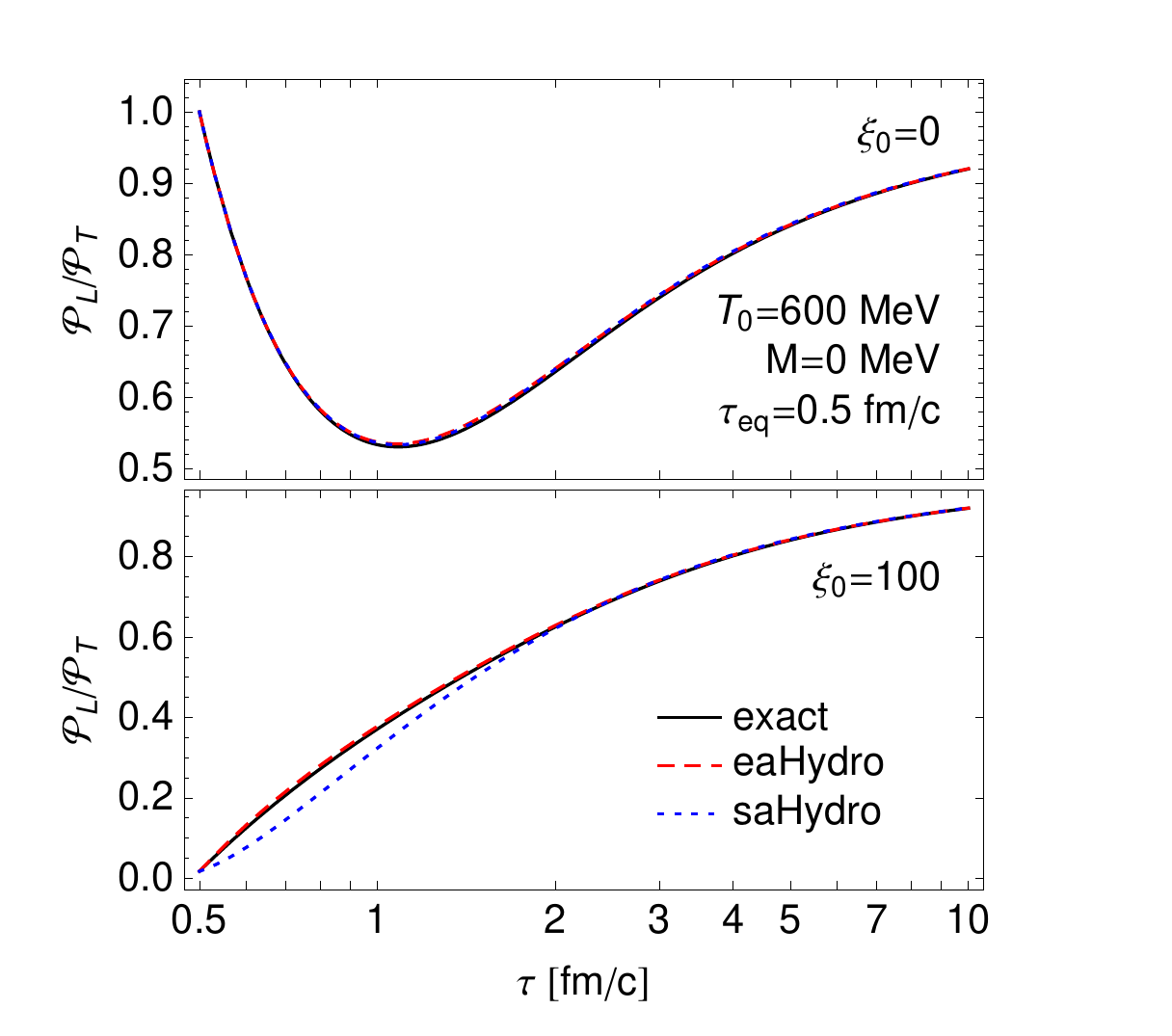}}
\caption{(Color online) Time dependence of the ratio of the longitudinal and transverse pressures.}
\label{fig:PLPTmass0}
\end{figure}

\begin{figure}[t]
\centerline{\includegraphics[angle=0,width=0.7\textwidth]{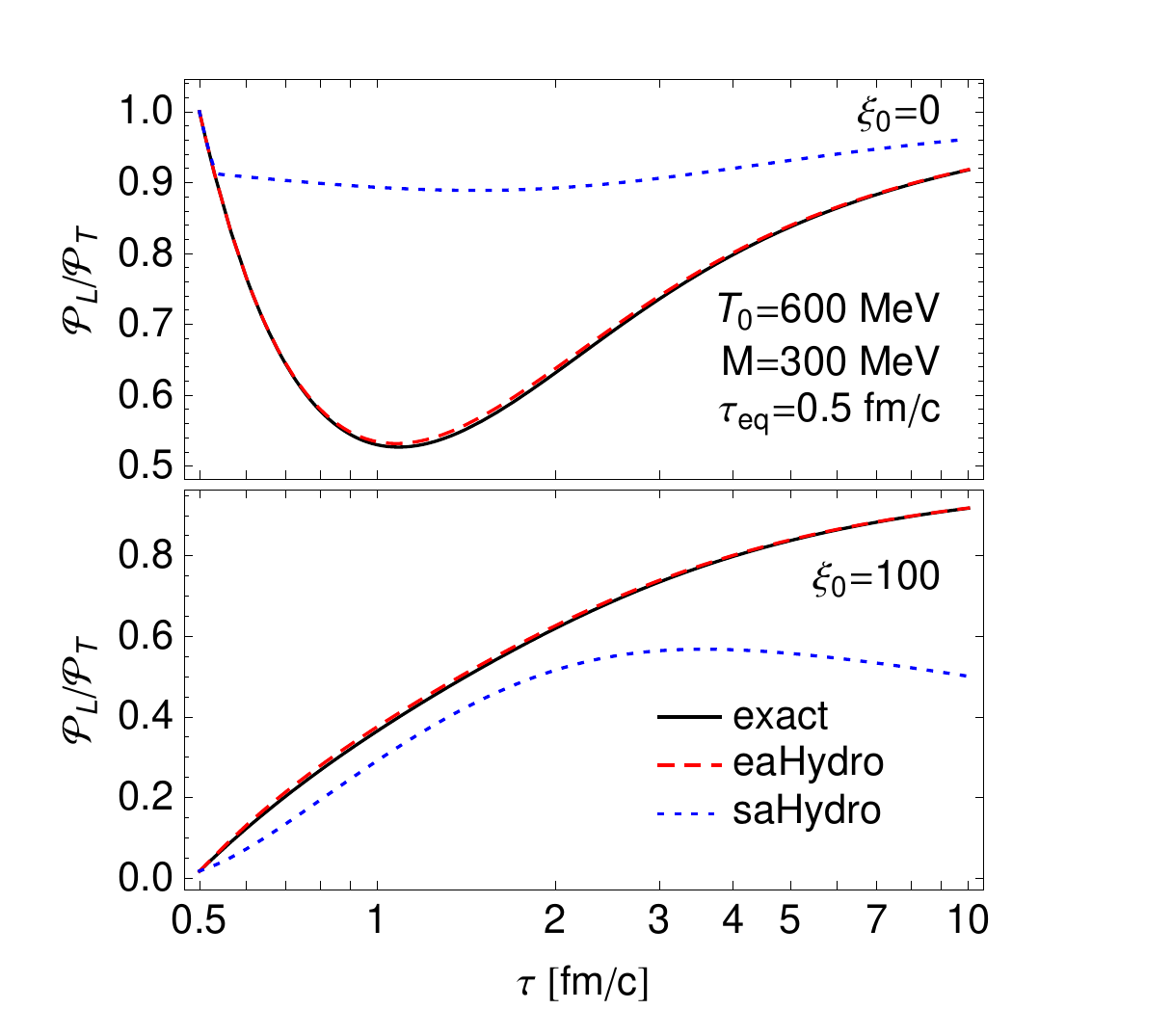}}
\caption{(Color online) Same as Fig.~\ref{fig:PLPTmass0} but now for the finite particle mass of 300 MeV.}
\label{fig:PLPTmass}
\end{figure}

In Fig.~\ref{fig:Tmass} we again show the time dependence of the effective temperature but now for the case of \mbox{$M=300$ MeV}. The notation is the same as in Fig.~\ref{fig:Tmass0}.  In this case we see small differences between the result of saHydro (dotted blue line) and the exact solution (solid black line).  On the other hand, the results of eaHydro (dashed red line) seem to agree quite well with the exact solution.

A more sensitive measure of the non-equilibrium dynamics of the system can be obtained by computing the ratio of the longitudinal and transverse pressures. In Fig.~\ref{fig:PLPTmass0} we show the time dependence of this ratio for the case $M=0$.  Once again we see that eaHydro agrees very well with the exact kinetic theory solution. The saHydro formulation on the other hand has relatively large deviations from the exact solution when the initial anisotropy of the system is large, see e.g. the bottom panel of Fig.~\ref{fig:PLPTmass0}.  

In Fig.~\ref{fig:PLPTmass} we show the ratio of the pressures for the case \mbox{$M=300$ MeV}. In this case saHydro shows large deviations from the exact solution.  In the top panel we see that the saHydro solution has a strong ``kink'' in the solution at early times.  We have investigated this and find that the saHydro equations become stiff when $M\neq0$ and the results obtained can be highly dependent on the parameters used and numerical algorithm employed.  The eaHydro approach which uses the second moment, does not seem to suffer from this problem.  Additionally, in the bottom panel of Fig.~\ref{fig:PLPTmass} we see another problem with the saHydro approach: for massive particles the system does not always approach isotropic equilibrium at late times.

\begin{figure}[t]
\centerline{\includegraphics[angle=0,width=0.7\textwidth]{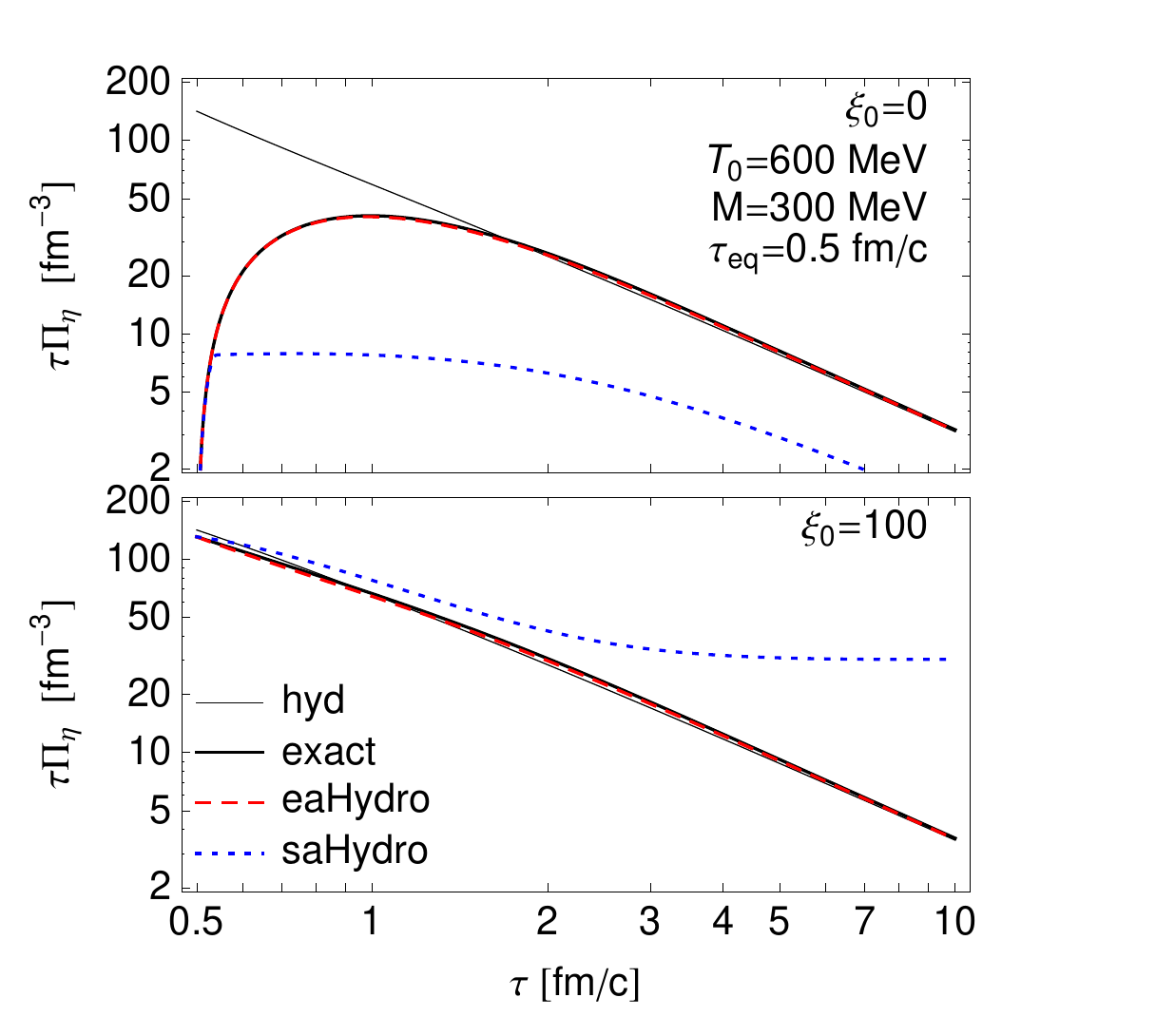}}
\caption{(Color online) Time dependence of the shear viscous pressure multiplied by $\tau$.}
\label{fig:shear}
\end{figure}

As an alternative measure of the pressure anisotropy, in Fig.~\ref{fig:shear} we show the time dependence of the shear viscous pressure multiplied by $\tau$ when \mbox{$M=300$ MeV}.  The shear viscous pressure is defined via
\begin{equation}
\Pi_\eta = \frac{2}{3} ( {\cal P}_T - {\cal P}_L ) \, .
\end{equation}
In Fig.~\ref{fig:shear} the thin black line labeled `hyd' is the first-order viscous hydrodynamics solution $\Pi_\eta^{\rm hyd} = 4 \eta/3\tau$ where, in relaxation time approximation, the shear viscosity is given by~\cite{Anderson:1974}
\begin{equation}
\eta(T) = \frac{\tau_{\rm eq} {\cal P}_{\rm eq}(T)}{15} \, \gamma^3\left[ \frac{3}{\gamma^2}\frac{K_3}{K_2}  -\frac{1}{\gamma}+\frac{K_1}{K_2}-\frac{K_{i,1}}{K_2} \right] ,
\label{etaAW}
\end{equation}
where all functions above are understood to be evaluated at $\gamma \equiv M/T$, $K_n$ are modified Bessel functions, and $2 K_{i,1} = \pi\left[1 - \gamma K_0(\gamma) L_{-1}(\gamma) - \gamma K_1(\gamma) L_{0}(\gamma) \right]$ where $L_i$ is a modified Struve function \cite{Florkowski:2014sfa}.  At late times we see that both the exact solution and the eaHydro approximation approach the first-order solution and are in good agreement with one another at all times.  The saHydro approach, however, has significant differences from the exact solution or exhibits unphysical behavior at late times.

\begin{figure}[t]
\centerline{\includegraphics[angle=0,width=0.7\textwidth]{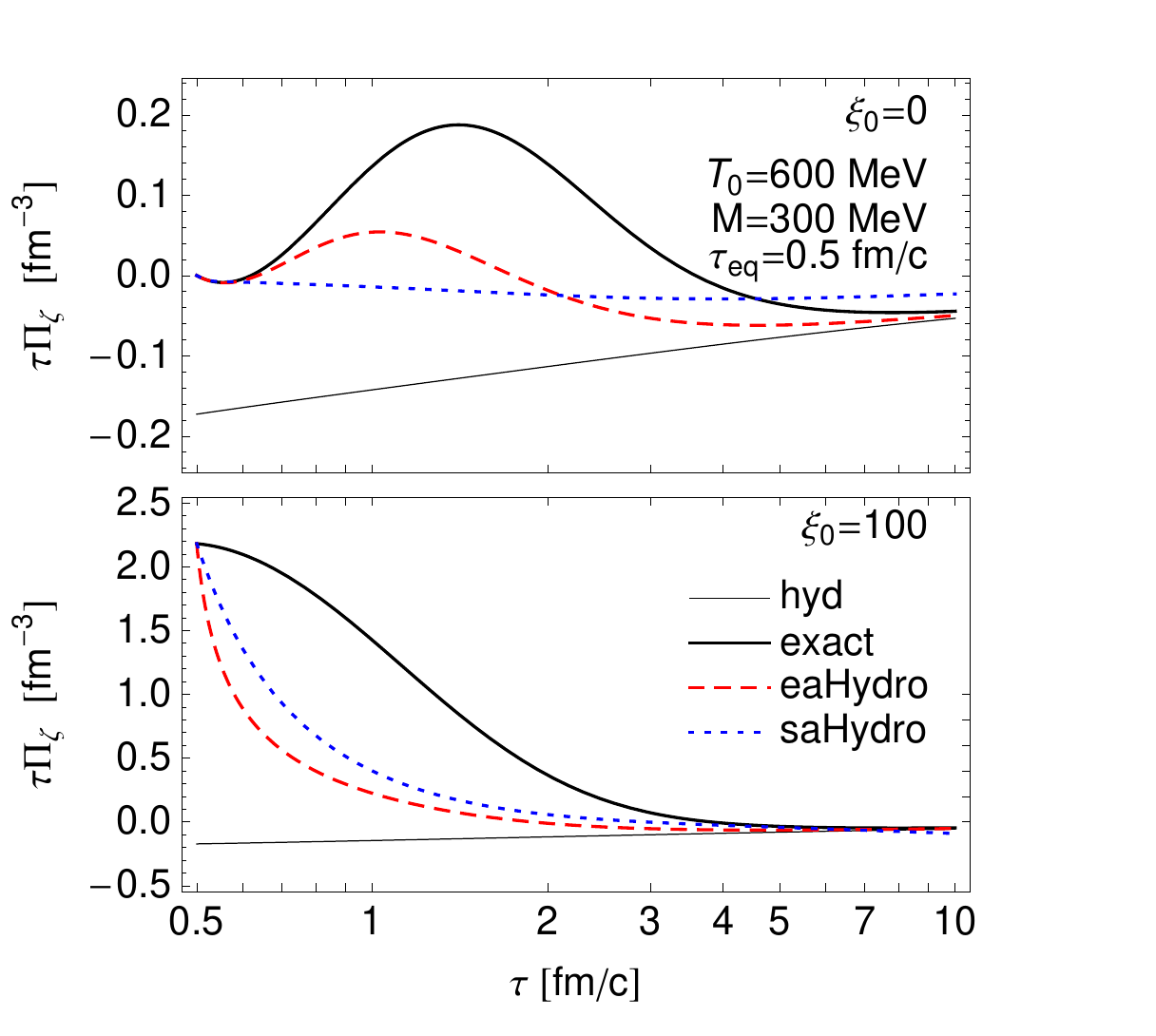}}
\caption{(Color online) Time dependence of the bulk viscous pressure multiplied by $\tau$.}
\label{fig:bulk}
\end{figure}

Finally, for $M\neq0$ the system will possess a non-vanishing bulk pressure
\begin{equation}
\Pi_\zeta(\tau) = \frac{1}{3}
\left[{\cal P}_L(\tau) + 2 {\cal P}_T(\tau)
- 3 {\cal P}_{\rm eq}(\tau) \right] .
\label{PIkz}
\end{equation}
In Fig.~\ref{fig:bulk} we show the time dependence of the bulk viscous pressure multiplied by $\tau$.
As before, we compare the exact solution (solid black line) with the eaHydro approach (red dashed line)
and the saHydro approach (blue dotted line) and the first-order viscous hydrodynamics solution (thin
black line labeled `hyd').  For the first-order solution one has $\Pi_\zeta^{\rm hyd} = - \zeta/\tau$
with $\zeta$ being the bulk viscosity which, for a massive Boltzmann gas in relaxation time approximation, 
is given by \cite{Bozek:2009dw,Sasaki:2008fg,Romatschke:2011qp,Florkowski:2014sfa}
\begin{equation}
\zeta(T) = \tau_{\rm eq} {\cal P}_{\rm eq} \, 
\frac{\gamma^2}{3} 
\left[-
\frac{\gamma K_2}{3 (3 K_3+\gamma K_2)}
+ \frac{\gamma}{3} \left(\frac{K_1}{K_2}-\frac{K_{i,1}}{K_2}\right) \right] .
\label{zetaPB}
\end{equation}
As we can see from this figure, neither of the aHydro approaches considered accurately describes
the evolution of the bulk pressure obtained from the exact kinetic solution.  We mention, however,
that this is somehow unsurprising since neither formalism has a parameter which plays the role of a bulk
viscosity in the microscopic form of the one-particle distribution function.  This suggests 
that one should extend aHydro to include the possibility of a bulk viscous term in the argument
of the distribution function.

\section{Conclusions}
\label{sect:concl}

In this paper we considered two methods for obtaining the anisotropic hydrodynamics equations of motion.
In the first approach (saHydro) we used the zeroth and first moments of the RTA kinetic equation and 
in the second approach (eaHydro) we used the first and second moments of the RTA kinetic equation.  We compared
the results of these two approximation schemes with a recent exact solution
of the one-dimensional boost-invariant dynamics in the relaxation time approximation \cite{Florkowski:2014sfa}.
In the massless case, both prescriptions agree quite nicely with the exact solution.
However, we found that the effective temperature and pressure anisotropy were most accurately described by
the eaHydro prescription which uses the second moment to obtain the necessary equation of motion 
\cite{Tinti:2013vba}, particularly when $M\neq0$.  For the bulk pressure neither approximation scheme
seems to accurately describe the bulk pressure evolution.\footnote{The accuracy compared to the 
exact bulk pressure solution is similar to second-order viscous hydrodynamic evolution of the bulk pressure 
\cite{Florkowski:2014sfa}.}

Looking to the future, these results tell us that one should use the second moment of the kinetic
equation to determine the evolution equation(s) for the anisotropy parameter(s). This 
relatively simple change could provide dramatic improvements to efforts underway to linearize
around anisotropic backgrounds~\cite{Bazow:2013ifa}.  As we have demonstrated herein, when one
uses the second moment to obtain the equations of motion, the resulting anisotropic hydrodynamics
framework (eaHydro) provides an excellent approximation to the exact solution.  As a result, 
linearizing around this background will further reduce the magnitude of higher-order corrections.

\acknowledgments

R.R. was supported by Polish National Science Center grant No. DEC-2012/07/D/ST2/02125, the 
Foundation for Polish Science, and U.S.~DOE Grant No. DE-SC0004104.   W.F. and L.T. were 
supported by Polish National Science Center grant No. DEC-2012/06/A/ST2/00390.  M.S. was 
supported in part by U.S.~DOE Grant No. DE-SC0004104.

\section*{Appendix: ${\cal H}_2$ functions}
\label{sect:app1}

In the Appendix we define the ${\cal H}_2$ functions which appear in the body of the text.  We
also list some useful properties of these functions that have been used to simplify the final 
equations of motion.

\subsection{Energy Density}
\label{sect:def1}

The energy density is expressed via the function
\begin{eqnarray}
\tilde{\cal H}_2(y,z) = \int\limits_0^\infty dr\, r^3 \, {\cal H}_2\left(y,\frac{z}{r} \right)
\, \exp\left(-\sqrt{r^2+z^2}\right) ,
\label{tildeH2}
\end{eqnarray}
where
\begin{eqnarray}
 {\cal H}_2(y,\zeta) &=& y \, 
 \int\limits_0^\pi d\phi \sin\phi
\, \sqrt{y^2 \cos^2\phi + \sin^2\phi + \zeta^2}
\nonumber \\
&=& y \left( \sqrt{y^2+\zeta^2} + \frac{1+\zeta^2}{\sqrt{y^2-1}}
\tanh^{-1} \sqrt{\frac{y^2-1}{y^2+\zeta^2}} \, \right).
\label{H2}
\end{eqnarray}

\subsection{Longitudinal Pressure}
\label{sect:def2}

The longitudinal pressure is defined by the function
\begin{eqnarray}
\tilde{\cal H}_{2L}(y,z) = \int\limits_0^\infty dr\, r^3 \, {\cal H}_{2L}
\left(y,\frac{z}{r} \right)
\, \exp\left(-\sqrt{r^2+z^2}\right),
\label{tildeH2L}
\end{eqnarray}
where
\begin{eqnarray}
\hspace{-1.75cm} {\cal H}_{2L}(y,\zeta) &=& y^3 \, 
 \int\limits_0^\pi \frac{d\phi \sin\phi \cos^2\phi }{\, \sqrt{y^2 \cos^2\phi + \sin^2\phi + \zeta^2}}
 \nonumber \\
\hspace{-1.75cm} &=& \frac{y^3}{(y^2-1)^{3/2}}
\left[
\sqrt{(y^2-1)(y^2+\zeta^2)}-(\zeta^2+1)
\tanh^{-1}\sqrt{\frac{y^2-1}{y^2+\zeta^2}} \,\,\right]. 
\label{H2L}
\end{eqnarray}
%

\subsection{Transverse Pressure}
\label{sect:def3}

The transverse pressure is expressed with the help of the function
\begin{eqnarray}
\tilde{\cal H}_{2T}(y,z) &=& \int\limits_0^\infty dr\, r^3 \, {\cal H}_{2T}
\left(y,\frac{z}{r} \right)
\, \exp\left(-\sqrt{r^2+z^2}\right),
\label{tildeH2T}
\end{eqnarray}
where
\begin{eqnarray}
{\cal H}_{2T}(y,\zeta) &=&
 y \, 
\int\limits_0^\pi \frac{d\phi \sin^3\phi }{
\, \sqrt{y^2 \cos^2\phi + \sin^2\phi + \zeta^2}} 
\label{H2T} \nonumber \\
&=& \frac{y}{(y^2-1)^{3/2}}
\left[\left(\zeta^2+2y^2-1\right) 
\tanh^{-1}\sqrt{\frac{y^2-1}{y^2+\zeta^2}}
-\sqrt{(y^2-1)(y^2+\zeta^2)} \right]. \hspace{1cm}
\end{eqnarray}
In our calculations, we insert the analytic expressions from Eqs.~(\ref{H2}), (\ref{H2L}), and (\ref{H2T}) into Eqs.~(\ref{tildeH2}), (\ref{tildeH2L}), and (\ref{tildeH2T}), respectively,  and numerically determine the values of the functions  $\tilde{\cal H}_2(y,z)$, $\tilde{\cal H}_{2L}(y,z)$, and $\tilde{\cal H}_{2T}(y,z)$. Besides the functions $\tilde{\cal H}_2(y,z)$, $\tilde{\cal H}_{2L}(y,z)$, and $\tilde{\cal H}_{2T}(y,z)$ it is also convenient to introduce the function
\begin{eqnarray}
\hat{\cal H}_2(y,z) = \int\limits_0^\infty dr\, r^3 \frac{z^2}{\sqrt{r^2+z^2}}\, {\cal H}_2\left(y,\frac{z}{r} \right)
\, \exp\left(-\sqrt{r^2+z^2}\right).
\label{hatH2}
\end{eqnarray}

\subsection{Derivatives}
\label{sect:der}

In the calculation of the time derivative of the energy density it is helpful to use the following identities
\begin{eqnarray}
\frac{\partial \tilde{\cal H}_2(y,z)}{\partial y}
= \frac{1}{y} \left[
\tilde{\cal H}_2(y,z)+\tilde{\cal H}_{2L}(y,z)
\right] ,
\label{dHdy}
\end{eqnarray}
and
\begin{eqnarray}
\frac{\partial \tilde{\cal H}_2(y,z)}{\partial z}
= \frac{1}{z} \left[
\tilde{\cal H}_2(y,z)-\tilde{\cal H}_{2L}(y,z)
-\tilde{\cal H}_{2T}(y,z)-\hat{\cal H}_2(y,z)
\right] .
\label{dHdz}
\end{eqnarray}
In the case where $y=(1+\xi)^{-1/2}$ and $z=M/\Lambda$, it is useful do define
\begin{eqnarray}
\Omega_2(\xi,\Lambda) =
\tilde{\cal H}_2((1+\xi)^{-1/2},M/\Lambda)+\tilde{\cal H}_{2L}((1+\xi)^{-1/2},M/\Lambda)  \, ,
\label{sigma2}
\end{eqnarray}
and
\begin{eqnarray}
\Omega_4(\xi,\Lambda) &=&
\tilde{\cal H}_2((1+\xi)^{-1/2},M/\Lambda)-
\tilde{\cal H}_{2L}((1+\xi)^{-1/2},M/\Lambda)
\nonumber \\
&& -
\tilde{\cal H}_{2T}((1+\xi)^{-1/2},M/\Lambda)
-
\hat{\cal H}_{2}((1+\xi)^{-1/2},M/\Lambda)  \, .
\end{eqnarray}
As a result, the total time derivative of $\tilde{\cal H}_2(y,z)$ can be expressed compactly as
\begin{eqnarray}
\frac{d}{d\tau} \tilde{\cal H}_2((1+\xi)^{-1/2},M/\Lambda)
=-\frac{\Omega_2(\xi,\Lambda)}{2(1+\xi)} \frac{d\xi}{d\tau} -\frac{\Omega_4(\xi,\Lambda)}{\Lambda}\frac{d\Lambda}{d\tau} \, .
\label{dH2dtau}
\end{eqnarray}

\bibliography{ahm}

\end{document}